\documentclass[prl,twocolumn,showpacs,amsmath,superscriptaddress]{revtex4}
\usepackage{graphicx}
\usepackage{multirow}

\newcommand{\BNRO}{Bi$_2$NiReO$_6$}
\newcommand{\BMRO}{Bi$_2$MnReO$_6$}
\newcommand{\BFCO}{Bi$_2$FeCrO$_6$}
\newcommand{\SCOO}{Sr$_2$CrOsO$_6$}


\begin{document}

% declarations for front matter
\title{High temperature multiferroicity and strong magnetocrystalline anisotropy in $3d-5d$ double perovskites}

\author{Marjana Le\v{z}ai\'{c}}\affiliation{Peter Gr\"{u}nberg Institut, Forschungszentrum
  J\"{u}lich, D-52425 J\"{u}lich and JARA-FIT, Germany}\email{m.lezaic@fz-juelich.de}\affiliation{Department of Materials, ETH Zurich, 
Wolfgang-Pauli Strasse 27, CH-8093 Zurich, Switzerland}
\author{Nicola A. Spaldin}\affiliation{Department of Materials, ETH Zurich, 
Wolfgang-Pauli Strasse 27, CH-8093 Zurich, Switzerland} 

%\date{\today}

\begin{abstract}
Using density functional calculations we explore the properties of
as-yet-unsynthesized $3d - 5d$ ordered double perovskites ($A_2BB'$O$_6$) with highly polarizable Bi$^{3+}$ ions on the
$A$ site. We find that the Bi$_2$NiReO$_6$ and Bi$_2$MnReO$_6$ compounds are insulating
and exhibit a robust net magnetization that persists above room temperature. When the in-plane lattice vectors of the pseudocubic unit cell are constrained to be orthogonal (for example, by coherent heteroepitaxy),
the ground states
are ferroelectric with large polarization and a very large uniaxial magnetocrystalline anisotropy with easy axis along the ferroelectric polarization direction. 
Our results suggest a route to multiferroism and electrically controlled magnetization orientation at room temperature.
\end{abstract}

\maketitle

\section{Introduction}

There is increasing current interest in developing multiferroic materials with ferromagnetic 
and ferroelectric order in the same phase for future spintronic or magnetoelectronic devices~\cite{Kleemann, Gajek, Chu, Lebeugle}.
Although new multiferroics are being predicted and synthesized at an accelerating pace, 
a major obstacle for their adoption in applications remains their low magnetic ordering temperatures,
which are
generally far below room temperature. The problem more generally lies in the scarcity of 
insulators with any net magnetization -- either ferro- or ferrimagnetic -- at room temperature, 
even before the additional requirement of polarization is included.
Most work on novel multiferroics has been restricted to $3d$ transition metal oxides
(see, for example, Refs. \cite{Baettig/Ederer/Spaldin:2005, Yamauchi, Picozzi, Giovanetti, Yamasaki, Sakai, Nechache, Hatt/Spaldin/Ederer:2010}), 
motivated by the expectation that $4d$ or $5d$ compounds
would likely be metallic. However, recent work on new $3d - 5d$ double perovskites
\cite{Krockenberger,Lee-Pickett} showed this to be a misconception, and identified a ferrimagnetic insulator,
\SCOO , with magnetic ordering temperature well above room temperature. 
First-principles calculations have been shown to accurately reproduce the measured magnetic
ordering temperatures,~\cite{Philipp, Kato} in the Sr$_2$Cr$M$O$_6$ series ($M$ = W, Re, Os), \cite{Mandal} and have been invaluable in explaining the origin of 
the robust ferrimagnetic ordering.~\cite{Sarma, Kanamori, Fang, Serrate}  

We build here on these recent developments to propose a route to achieving multiferroics
with higher magnetic ordering temperatures. We use $3d - 5d$ double perovskites to achieve
high temperature magnetism combined with insulating behavior, and introduce lone-pair active cations 
on the A sites to induce polarizability~\cite{Spaldin-Rabe:1999, Seshadri/Hill:2001}. 
The heavy Bi and $5d$ elements provide an additional desirable feature: The spin-orbit interaction strongly couples the magnetic easy axis to the ferroelectric polarization direction.
Specifically, we explore two compounds, \BMRO\ and
\BNRO. We use density functional theory to calculate their zero Kelvin
structure and magnetic ordering, and Monte Carlo simulations with parameters extracted
from the first-principles calculations to calculate their magnetic ordering temperatures. We
find that both materials are magnetic insulators, with ordering temperatures well above room 
temperature. While the global ground state in both materials is anti-polar, we show that
when the in-plane lattice vectors form an angle close to 90$^\circ$  
a ferroelectric phase results. In practice this could be achieved using coherent
heteroepitaxy and the resulting ferroelectricity could be manipulated using strain.  
We show that the magnetic behavior results from an antiferromagnetic 
superexchange~\cite{Anderson, Mandal}
and the ferroelectricity from the Bi$^{3+}$ lone pairs, reminiscent of other Bi-based multiferroics. Finally, we
investigate the effect of strong spin-orbit coupling (SOC) due to the presence of heavy 
Bi$^{3+}$ and Re$^{4+}$ cations on the structural, magnetic, and ferroelectric properties of these compounds.

\section{Method of calculation}

Structural optimizations were performed using the Vienna {\it Ab-initio} Simulation Package 
(VASP) with projector augmented wave (PAW) potentials;~\cite{VASP} the semicore $p$ states of Mn, Ni, and Re and the $d$ states of 
Bi were included in the valence. We used an energy cutoff of 500 eV
and $6\times6\times6$, $4\times4\times6$, and $4\times4\times4$ $\Gamma$-point centered
$k$-point meshes for unit cells containing 10, 20, and 40 atoms
respectively. The exchange-correlation functional was treated within the generalized gradient approximation
(GGA);~\cite{Perdew} while extension to the GGA$+U$ method had only a small influence on the electronic 
properties, we discuss later its effect on the structural behavior. 
Magnetic ordering temperatures were obtained using a finite-temperature Monte Carlo scheme
within a Heisenberg Hamiltonian $H = -\frac{1}{2}\sum_{i,j} J_{i,j} {\bf M}_{i}\cdot{\bf M}_{j}$, where 
${\bf M}_{i}$ and ${\bf M}_{j}$ are the magnetic moments on sites $i$ and $j$ of the crystal lattice. 
The exchange constants $J_{i,j}$ were calculated in the frozen-magnon scheme~\cite{Lezaic} using
the all-electron full-potential linearized augmented plane-wave code {\tt FLEUR}~\cite{FLEUR}, 
with a plane-wave cutoff of 4.2 hartrees and $15\times15\times15$ $k$-points and a $6\times6\times6$ 
spin-spiral grid in a 
10-atom unit cell. Ferroelectric polarizations were extracted from the shifts of the centers 
of Wannier functions~\cite{Freimuth}. The magnetocrystalline anisotropy energy was extracted from the self-consistent total-energy calculations within the {\tt FLEUR} code, on a $13\times13\times13$ $k$-point grid.

First, we calculate the lowest energy structure for both \BMRO\ and \BNRO\ within the
GGA approximation. We proceed by successively freezing in the 12 centrosymmetric combinations of
tilts and rotations of the oxygen octahedra that can occur in ordered double
perovskites~\cite{Howard}. We then further reduce the symmetry of these lowest energy tilt systems
by displacing the anions and cations relative to each other in the manner of a polar
displacement. 
Next we optimize the atomic positions (by minimizing the Hellmann-Feynman forces), as well as 
the unit cell shape and volume within each symmetry. 
In this first series of optimizations we assume ferrimagnetic ordering and do not include 
SOC; the influence of SOC on the 
lowest-energy structures is examined later. 

\section{Ground-state structure and influence of epitaxial constraints}
\begin{figure}
\begin{center}
(a) \includegraphics*[width=0.56\linewidth]{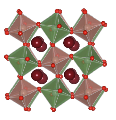}\hspace{0.02\linewidth}\\
(b)  \includegraphics*[width=0.56\linewidth]{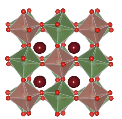}
\caption{(Color online) (a) Ground-state $P2_1/n$ structure and (b) the ferroelectric $R3$ structure of the double perovskites \BMRO\ and \BNRO. Bi$^{3+}$ cations are depicted as the large (brown) spheres and O$^{2-}$ as the small (red) spheres. Re$^{4+}$ and Mn$^{2+}$ (Ni$^{2+}$) are alternating in the shaded octahedra, in a three-dimensional checkerboard manner (this is indicated by the color of the octahedra). While in the $P2_1/n$ symmetry the Bi$^{3+}$ cations form an ``antipolar'' pattern, their coherent shift along the [111] direction can be clearly seen in the  $R3$ structure.\label{fig:structure}}
\end{center}
\end{figure}

Our calculations yield the same monoclinic, centrosymmetric  $P2_1/n$ symmetry ground state for both compounds. It is characterized by an $a^-a^-c^+$ tilt pattern of the oxygen 
octahedra. Two lattice vectors are identical with angles of 93.6$^{\circ}$ and 93.5$^{\circ}$
between them in the Mn and Ni compounds, respectively (in the following we refer to these as
the ``in-plane'' lattice vectors); the third lattice vector is perpendicular to 
the first two and of different length. 
Importantly, in both compounds there is a slightly higher energy ferroelectric (FE) structure 
with rhombohedral $R3$ symmetry (see Table~\ref{tab:energies}). 
\begin{table}[b]

   \begin{tabular*}{0.47\textwidth}{c|c|@{\extracolsep{\fill}}c c}
    Symmetry&Tilt system & \multicolumn{2}{c}{Total energy}\\   &(in Glazer notation~\cite{Glazer}) &\BMRO\ &\BNRO\\       
    \hline
    $Fm\bar 3m$& $a^0a^0a^0$ & 1.86 eV &  2.02 eV\\ 
    $R\bar3$& $a^-a^-a^-$ & 248 meV & 333 meV \\
    $R3$& $a^-a^-a^-$ + FE shift & 32 meV & 18 meV \\
    \hline
    
    \end{tabular*} \caption{Total energies (per formula unit) of the lowest-energy phases of \BMRO\ and \BNRO\  and the cubic $Fm\bar 3m$ phase with respect to their ground state $P2_1/n$ phase.  }\label{tab:energies}
\end{table}
The $R3$ structure corresponds to an $a^-a^-a^-$ tilt pattern 
yielding lattice vectors of equal length and rhombohedral angles of 60$^{\circ}$ and 61$^{\circ}$
for the Mn and Ni compounds respectively, 
combined with polar relative
displacements of anions and cations along the [111] rhombohedral axis induced by the 
well-established stereochemically active Bi lone pair~\cite{Spaldin-Rabe:1999} (see Fig.~\ref{fig:structure}). 
The energy of the FE phase is significantly lower than that of the corresponding
centrosymmetric $R\bar3$ phase, suggesting a high ferroelectric ordering temperature if
this symmetry could be stabilized. For comparison, in the prototypical multiferroic BiFeO$_3$ with $T_{\rm C}$ of 1103~K, we obtain a difference of  515 meV per two formula units between the corresponding states. 
The calculated polarization in both cases is along the [111] direction, and comparable in size to that of BiFeO$_3$ (see Table~\ref{tab:properties}). 

\begin{figure*}
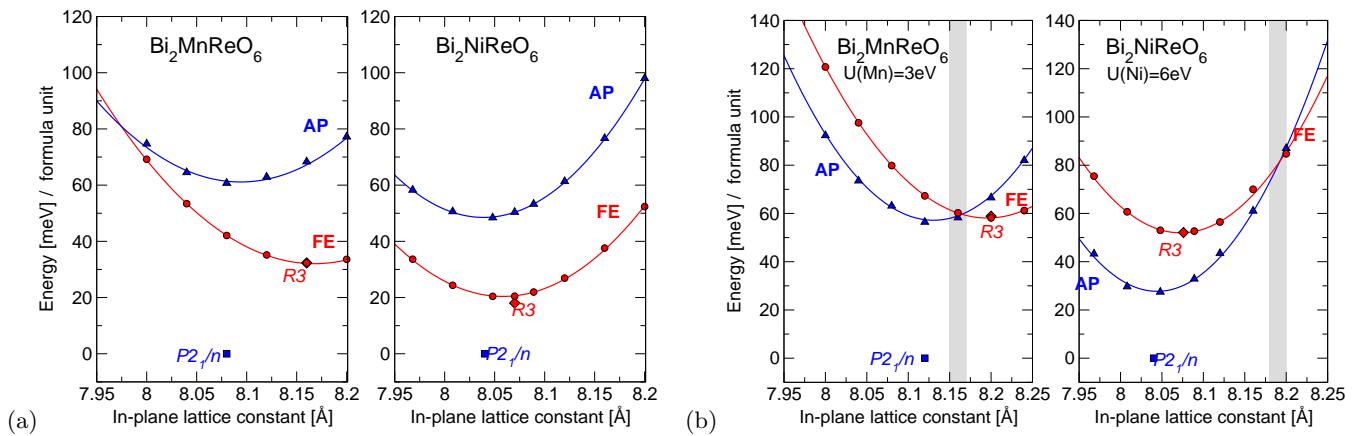

\begin{center}
(a)\includegraphics*[width=0.46\linewidth]{noU.eps}\hspace{0.02\linewidth}(b)  \includegraphics*[width=0.46\linewidth]{withU.eps}
\caption{(Color online) (a) Distorted \BMRO\ and \BNRO\ obtained from the $R3$ (red circles) and $P2_1/n$ (blue 
triangles) structure by constraining the two in-plane lattice vectors to be equal and the
angle between them to 90$^\circ$. The out-of-plane parameter and the ionic positions were relaxed. The energies are positioned with respect to the bulk 
unconstrained $P2_1/n$ ground state (blue squares). The red diamonds show the energies for the unconstrained  $R3$ structure. A crossover between the antipolar (AP) and the ferroelectric (FE) state 
occurs at 1.8\% (with respect to the minimum of the FE curve) of compressive strain in \BMRO. Note that the energy of the $R3$ state lies on the strain curve for 
\BMRO, but is somewhat lower for  \BNRO, due to the choice of a square in-plane lattice: the rhombohedral 
angle in $R3$ phase of \BMRO\ is 60$^\circ$, while in \BNRO\ it is 61$^\circ$.
(b) Similar to (a), with an addition of a Hubbard $U$ on Mn (3~eV) and Ni (6~eV) $3d$ states.  The shaded area shows the AP-FE crossover region. \label{fig:strain}}
\end{center}
\end{figure*}

While the ground state $P2_1/n$ phase is of interest in its own right as a possible 
high temperature magnetic insulator, we next explore whether it is possible to
identify conditions that
stabilize the $R3$ ferroelectric phase. We use the fact that in the FE $R3$ phase 
all three lattice vectors have equal length, and the rhombohedral angle is equal
or close to the ideal value of 60$^{\circ}$. In contrast, in the $P2_1/n$ phase,
only the lengths of the two in-plane lattice vectors are equal, and the inter-in-plane 
angles deviate strongly from the ideal 90$^{\circ}$. Motivated by these observations,
we first constrain all three lattice parameters to be equal in length and indeed find that this
constraint stabilizes the FE phase. Next we enforce that only two of the three lattice 
parameters are equal in length, but constrain them to be perpendicular to each other 
and again find that the FE phase is the lowest energy
\footnote{
Note that the lowest energy $P2_1/n$ structure would not be readily identified as the 
ground state through a search for unstable phonons starting from the prototype cubic 
perovskite structure because it only becomes lowest in energy when the lattice parameters 
are allowed to be unequal. Indeed, when we perform full structural optimizations within
the LDA+U method for
the previously studied double perovskite, \BFCO~\cite{Baettig}, we find the $P2_1/n$
structure to be lower in energy than the previously reported $R3$ when we allow for
lattice parameters of different length. Interestingly, in this case, the GGA+U 
method yields 
a ferroelectric ground state, consistent with some experimental reports~\cite{Nechache}. }.
Such a constraint could be achieved in practice through coherent heteroepitaxy between 
a thin film and a substrate, and is often described as a biaxial strain state in the
literature.
In Fig.~\ref{fig:strain}(a), we show the calculated energies of the previously 
$R3$ and $P2_1/n$ structure types subject to this additional constraint, with the
in-plane lattice parameters varied over a range of realistic substrate strains 
corresponding to (001) epitaxial growth.
For each in-plane lattice parameter we allow the length and angle of the out-of-plane 
lattice parameter and atomic positions to relax to their lowest-energy configurations.
Our main finding is that, under the investigated constraint,
the FE phase is lower in energy than the antipolar (AP) phase. At the
optimized value of the in-plane lattice parameter (8.16~\AA\ for \BMRO\ and 8.07~\AA\  for \BNRO)
the energy differences between the FE and AP phases are 36 and 30~meV per formula unit, 
respectively. 
The FE phases are robustly insulating with calculated band gaps of 0.7 and 0.3~eV, respectively,
and have large polarizations of 84 $\mu$C/cm$^2$ 
(\BMRO) and 81 $\mu$C/cm$^2$ (\BNRO) along the pseudocubic [111] direction. Notice, however, that as in-plane compressive strain is applied, the differences in energies 
reduce, and in fact in \BMRO\ a crossover is reached at a compressive strain value of 1.8\%.
We therefore expect that both compounds, although AP in bulk, will in fact be ferroelectric 
and hence multiferroic in thin-film form, over a range of experimentally accessible strains. We 
see later that SOC tips the scale toward the FE phase, while electron correlations favor the AP phase.

\section{Electronic and magnetic properties of the ferroelectric phase}

\begin{figure}
\begin{center}
\includegraphics*[width=0.775\linewidth]{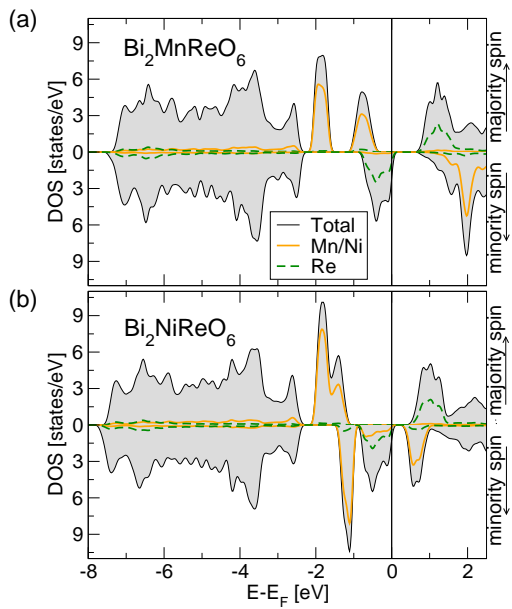}
\caption{(Color online) Density of states (DOS) of the ferroelectric phase of \BMRO\ (a) and \BNRO\ 
(b). The local magnetic moments on Mn/Ni and Re are antialigned. Thin black 
line, total DOS; green dashed line, Re; orange solid line, Mn/Ni. \label{fig:DOS}}
\end{center}
\end{figure}

Next, we analyze the magnetic behavior of the two ferroelectric compounds. We find that in
both cases the lowest-energy ordering is ferrimagnetic, with the moments on the $3d$ and
$5d$ transition metal ions antialigned (see Fig.~\ref{fig:DOS}). In both compounds, 
Re is in formal oxidation state 4$^+$  
corresponding to a filled $d-t_{2g}$ manifold in the minority spin-channel.
The oxidation states of Mn in \BMRO\ and of Ni in 
\BNRO\ are 2$^+$: $t_{2g}^3$~$e_g^2$ (high spin) for 
Mn and $t_{2g}^6$~$e_g^2$ for Ni. Due to these specific configurations of the outer electronic shells, and the nearly 150$^\circ$ Mn-O-Re (Ni-O-Re) angle, the 
magnetic moments on Re and Mn/Ni are coupled via an antiferromagnetic superexchange mechanism.~\cite{Goodenough, Kanamori:1959, Wollan}
The result is a ferrimagnetic configuration with a total spin moment of 2~$\mu_B$ in \BMRO\ and 1~$\mu_B$ in \BNRO. 

Our calculated magnetic ordering temperatures in the ferroelectric phase are 330~K for \BMRO\ and 
360~K for \BNRO, both significantly above room temperature. As a comparison, using the same method, 
we calculate a magnetic ordering temperature of 670~K for \SCOO\  (the experimental value is 725~K). Note that if growth can only be achieved in the form of ultra-thin films, the ordering temperatures will likely be reduced.

\section{Spin-orbit coupling and correlations effects}

Since the heavy Re and Bi atoms in our double perovskites are also likely to exhibit strong SOC, and consequently significant magnetostructural coupling,~\cite{Serrate} we repeat
our calculations with SOC explicitly included. We find that the ground state remains 
$P2_1/n$ AP, but now its energy difference from the FE state with $R3$ symmetry is reduced to 10~meV in
 \BMRO\ and only 2~meV in \BNRO. Both compounds still have semiconducting gaps which are now somewhat reduced while the ferroelectric polarization of the $R3$ phase is slightly increased (see Table~\ref{tab:properties}). The total magnetic moments (spin~+~orbital) 
amount to 2.34 and 0.58~$\mu_B$, respectively. Note that, in contrast to half-metallic ferromagnets where the SOC introduces states in the minority-spin gap yielding a noninteger total magnetic moment, here the noninteger moment is the consequence of the mixing of spin-up and spin-down states (spin is no longer a good quantum number with SOC included), while the gap is preserved.

\begin{table}

   \begin{tabular*}{0.47\textwidth}{c|c|c|c|c}
 SOC & Compound & Band gap [eV]& $P$  ($\mu$C/cm$^2$) & $M$ ($\mu_B$/f.u.)\\ 
 \hline
  \multirow{2}{*}{Off}& \BMRO & 0.7 & 84 & 2\\ 
   & \BNRO & 0.3 & 78 & 1 \\
   \hline
    \multirow{2}{*}{On}& \BMRO & 0.4 & 86 & 2.34\\ 
   & \BNRO & 0.2 & 80& 0.58 \\
    \hline
    \end{tabular*} \caption{Properties of the bulk ferroelectric $R3$ phase in \BMRO\ and \BNRO\ with and without the SOC included: band gap, ferroelectric polarization $P$ and the total magnetic moment $M$ (spin+orbital in case of SOC) per formula unit (f.u.). }\label{tab:properties}
\end{table}
In both compounds the magnetic easy axis lies along the ferroelectric polarization direction. The associated anisotropy energy is large and comparable to that in current magnetic recording media:
it amounts to 7 and 5.5 meV per formula unit in \BMRO\ and \BNRO, respectively. This suggests an exciting possibility of electrical control of the magnetization direction.

Finally, we investigate the influence of electronic correlations by adding a Hubbard $U$ on the $d$ states of Mn and Ni. 
In order to reduce the computational effort, we do not include the SOC. We keep in 
mind, however, that a consequence of its inclusion, as we have seen, is a reduction in the energy of 
the FE state with respect to the AP one.
Introduction of Hubbard $U$ corrections 
(on Mn $U$=3~eV and $J$=0.87~eV,  and on Ni $U$=6~eV and $J$=0.90~eV) 
does not change the ground state structure ($P2_1/n$ still has the lowest energy). Interesting quantitative
changes occur in the strain dependence however [Fig.~\ref{fig:strain}(b)]. 
While within the GGA the ferroelectricity was robust to the choice of substrate lattice
parameters over a likely range of accessible strains provided that the in-plane lattice 
parameters were constrained to be perpendicular to each other, in GGA$+U$ we find a cross-over between the FE
and AP states at moderate strain values. 
This suggests the intriguing possibility that an external strain, for example from a piezoelectric
substrate~\cite{Rata} could induce a ``dipole-flop'' transition
from a non-polar structure in which the dipoles are antialigned to a ferroelectric one. For the assumed 
value of $U$, the crossover region [shaded area in Fig~\ref{fig:strain}(b)] in \BMRO\ is within less than 1\% 
from either of the two minima: a small compressive strain will push the system toward the AP state, while a 
small tensile strain will favor a FE state. In \BNRO\ the applied $U$ pushes the AP state lower in energy, but the
crossover region to the FE state is still within an experimentally accessible range of 1.5\% tensile strain. While we can not make a quantitative prediction, 
since the relative energies of the AP and the FE state depend 
strongly on the value of the applied $U$ correction, 
within a reasonable range of $U$ the crossover region lies within 
2\% strain from the ground state structure.  

\section{Conclusion}

In summary, we predict from first-principles calculations that the $3d-5d$ double-perovskite compounds
\BMRO\ and \BNRO\ are high-T$_{\rm C}$ ferrimagnetic insulators. Moreover, it is likely possible 
to stabilize them in thin-film form in a ferroelectric phase with high polarization along the [111] 
direction. The magnetic easy axis lies along the ferroelectric polarization direction with a very large associated anisotropy energy. The different sizes and charges of the $3d$ and $5d$ transition metal ions suggest that
the required $B$-site ordering should be experimentally feasible; we hope that our findings will 
initiate experimental efforts to synthesize these novel multiferroic compounds.
 
%\acknowledgments
We thank Drs.\ Kris Delaney, Phivos Mavropoulos, Stefan Bl\"ugel, Frank Freimuth and Sergey Ivanov for many valuable discussions.
M.L. gratefully acknowledges the support of Deutsche Forschungsgemeinschaft, grant LE 2504/1-1, and the Young Investigators Group Programme of Helmholtz Association, contract VH-NG-409, as well as the J\"ulich Supercomputing Centre.
NS acknowedges support from the NSF NIRT program, grant number 0609377.

\end{document}